\documentclass[
 reprint,
 superscriptaddress,
 amsmath,amssymb,
 aps,
 pra,
 floatfix,
]{revtex4-2}
\pdfoutput=1

\usepackage{dcolumn}
\usepackage{bm}
\usepackage[colorlinks=true, linkcolor=black, urlcolor=black, citecolor=black]{hyperref}
\usepackage[normalem]{ulem}
\usepackage{bibunits}
\defaultbibliography{references}
\defaultbibliographystyle{apsrev4-1}
\usepackage{stfloats}

\usepackage{tikz}
\usetikzlibrary{quantikz2}
\usepackage[T1]{fontenc}
\usepackage{xcolor}
\definecolor{darkblue}{rgb}{0,0,0.5}
\definecolor{darkgreen}{rgb}{0,0.5,0}
\definecolor{darkred}{rgb}{.7,0,0}
\definecolor{purple}{rgb}{0.5,0,0.6}
\definecolor{orange}{rgb}{1,0.5,0}
\definecolor{grey}{rgb}{.6,.6,.6}
\definecolor{lightpink}{rgb}{1,0.7,0.75}
\definecolor{pink}{rgb}{1,0.4,0.58}
\definecolor{deeppink}{rgb}{1,0.08,0.58}
\definecolor{brown}{rgb}{0.59, 0.29, 0.0}
\definecolor{blue-green}{rgb}{0.0, 0.87, 0.87}

\usepackage{amsmath}
\usepackage{graphicx}
\usepackage{xcolor}

\newcommand{\sqc}{Silicon Quantum Computing Pty Ltd, UNSW Sydney, Sydney, New South Wales, Australia}

\usepackage[british]{babel}

\begin{document}

\title{
An 11-qubit atom processor in silicon
}

\author{Hermann Edlbauer}
    \affiliation{\sqc}
    \affiliation{\rm These authors contributed equally to this work.}
\author{Junliang Wang}
    \affiliation{\sqc}
    \affiliation{\rm These authors contributed equally to this work.}
\author{A. M. Saffat-Ee Huq}
    \affiliation{\sqc}
\author{Ian Thorvaldson}
    \affiliation{\sqc}
\author{Michael T. Jones}
    \affiliation{\sqc}
\author{Saiful Haque Misha}
    \affiliation{\sqc}
\author{William J. Pappas}
    \affiliation{\sqc}
\author{Christian M. Moehle}
    \affiliation{\sqc}
\author{Yu‐Ling Hsueh}
    \affiliation{\sqc}
\author{Henric Bornemann}
    \affiliation{\sqc}
\author{Samuel K. Gorman}
    \affiliation{\sqc}
\author{Yousun Chung}
    \affiliation{\sqc}
\author{Joris G. Keizer}
    \affiliation{\sqc}
\author{Ludwik Kranz}
    \affiliation{\sqc}
    \affiliation{\rm These authors jointly supervised this work.}
\author{Michelle Y. Simmons}
    \affiliation{\sqc}
    \affiliation{\rm These authors jointly supervised this work.}
    \affiliation{\rm Corresponding author: \href{mailto: michelle.simmons@sqc.com.au}{michelle.simmons@sqc.com.au}}

\date{\today}

\begin{abstract}
    Phosphorus atoms in silicon are an outstanding platform for quantum computing as their nuclear spins exhibit coherence time over seconds~\cite{Kane1998,stano2022review}.
By placing multiple phosphorus atoms within a radius of a few nanometers, they couple via the hyperfine interaction to a single, shared electron.
Such a nuclear spin register enables multi-qubit control above the fault-tolerant threshold~\cite{Madzik2022} and the execution of small-scale quantum algorithms~\cite{Thorvaldson2024}.
To achieve quantum error correction, fast and efficient interconnects have to be implemented between spin registers while maintaining high fidelity across all qubit metrics.
Here, we demonstrate such integration with a fully controlled 11-qubit atom processor composed of two multi-nuclear spin registers which are linked via electron exchange interaction.
Through the development of scalable calibration and control protocols, we achieve coherent coupling between nuclear spins using a combination of single- and multi-qubit gates with all fidelities ranging from 99.5\% to 99.99\%.
We verify the efficient all-to-all connectivity by preparing both local and non-local Bell states with a record state fidelity beyond 99\% and extend entanglement through the generation of Greenberger-Horne-Zeilinger (GHZ) states over all data qubits.
By establishing high-fidelity operation across interconnected nuclear-spin registers, we realise a key milestone towards fault-tolerant quantum computation with atom processors.
\end{abstract}

\maketitle

\section*{Introduction}

The predominant material in modern classical computers, silicon, is also a strong contender for the practical implementation of quantum processors~\cite{Burkard2023,stano2022review,Takeda2022,neyens2024probing}.
To unlock the promised computational benefits of quantum computing, the qubit count needs to scale while maintaining high operation fidelity and connectivity.
In terms of qubit numbers, the lead is currently held by superconducting~\cite{Google2024,Arute2019}, ion-trap~\cite{paetznick2024demonstration}, and neutral-atom~\cite{Bluvstein2023} processors, which approach hundreds of interconnected qubits.
Further scale-up faces platform-specific challenges related to manufacturing, control-systems miniaturisation, and materials engineering.
In this context, silicon quantum processors are emerging as a promising platform owing to their small footprint and materials compatibility with industrial manufacturing~\cite{zwerver2022qubits, neyens2024probing, Intel2024arxiv}.

In semiconductor devices, the number of individual qubits is rising with gate-defined arrays hosting up to 16 quantum dots~\cite{Intel2024arxiv,borsoi2024shared}.
To date, however, no more than four interconnected spins were used in the execution of quantum circuits owing to challenges associated with multi-qubit control~\cite{Hendrickx2021,Philips2022,Zhang2024}.
In this context, quantum computing with precision-placed phosphorus atoms in silicon, which we refer to as the 14|15 platform (according to the respective positions in the periodic table), is attracting growing interest driven by industry-leading physical-level metrics~\cite{stano2022review} with exceptional, second-long coherence times~\cite{Muhonen2014,hsueh2023}.
The 14|15 platform uses precision manufacturing~\cite{Fuechsle2012} to place individual phosphorus atoms in close proximity  ($\lesssim3$ nm) to each other in which a single loaded electron exhibits a hyperfine interaction with multiple nuclei.
Such spin registers provide a unique set of advantages: the shared electron naturally acts as an ancilla qubit enabling quantum non-demolition (QND) readout of the nuclear spins and native multi-qubit (Toffoli) gates~\cite{Madzik2022, Thorvaldson2024}.
Combined with recent advances in silicon purification with sub-200ppm of $^{29}$Si~\cite{Reiner2024}, these features enabled nuclear-nuclear CZ operations with fidelities exceeding 99\% and the execution of three-qubit algorithms on a single multi-spin register~\cite{Thorvaldson2024}.

To enable the scaling of the 14|15 platform, it is essential to develop fast interconnects between quantum processing nodes where each operates above the fault-tolerant threshold~\cite{vandersypen2017interfacing}.
The coupling of spin qubits is achievable by various mechanisms such as dipolar interaction~\cite{Sarkar2022} or spin-photon conversion in superconducting cavities~\cite{Dijkema2024}.
The fastest coupling mechanism is provided by exchange interaction, as demonstrated with a 0.8-ns $\sqrt{\rm SWAP}$ gate between atomic qubits in natural silicon~\cite{He2019}.
Exchange gates on electron spins have also been implemented with gate-defined quantum-dots in isotopically pure silicon with fidelities >99\%~\cite{Noiri2022, Mills2022,Xue2022,Wu2024}.
For atom qubits, the implementation of exchange gates has been attempted in purified silicon-28~\cite{stemp2024tomography}, but performance above the fault-tolerance threshold
has not been achieved so far.

Here, we report a precision-placed 11-qubit atom processor in isotopically purified silicon-28 that runs on a fast and efficient exchange-based link.
Compared to the previous atom-based implementations with nuclear spin qubits~\cite{Madzik2022, Reiner2024, Thorvaldson2024}, we triple the number of fully-interconnected data qubits while maintaining high performance well above 99\% fidelity.
This achievement is enabled by systematic investigations of qubit stability, contextual errors, and crosstalk, which informed the development of scalable calibration and control protocols.
After outlining the basic setup of the 11-qubit atom processor, we report single- and two-qubit metrics and show efficient all-to-all connectivity across each nuclear spin before finally demonstrating full entanglement of 8 data qubits.
Our results demonstrate that excellent qubit metrics are maintained when interconnecting quantum-computing nodes\textemdash an essential ingredient for quantum error correction.

\section*{Results}

\begin{figure*}[p!]
\centering
\includegraphics{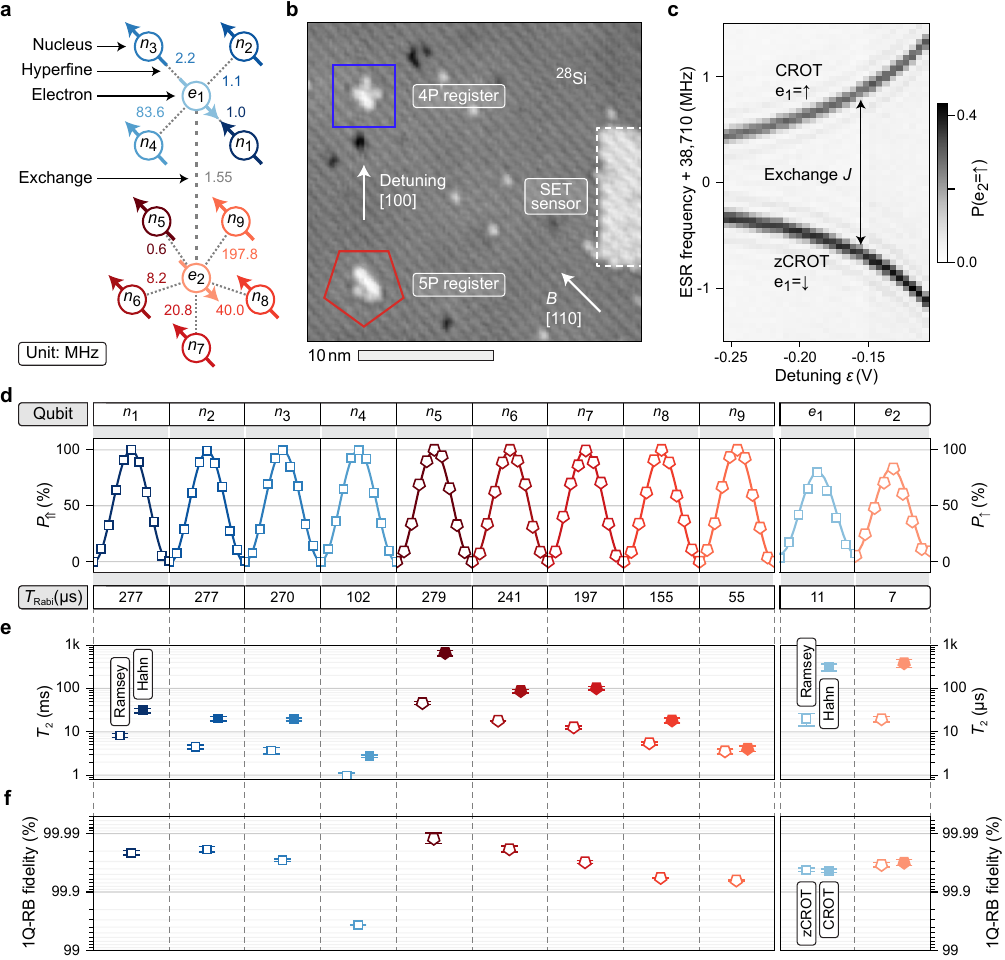}	
\caption{
\textbf{Single-qubit characteristics of the 11-qubit atom processor.}
\textbf{a,}
Connectivity of nuclear spins ($n_1$-$n_9$) and electron spins ($e_1$ and $e_2$) via hyperfine- and exchange-coupling with energies in MHz.
\textbf{b,}
Scanning tunneling micrograph of the processor core after hydrogen lithography showing the 4P register hosting $n_1$-$n_4$ and $e_1$ (square) and the 5P register hosting $n_5$-$n_9$ and $e_2$ (pentagon).
The distance of 13(1)~nm (center-to-center) between the nuclear spin registers is atomically engineered to enable exchange coupling~\cite{He2019,Kranz2023}.
\textbf{c,}
Exchange ESR spectrum of $e_{2}$ as function of voltage detuning $\varepsilon$. 
\textbf{d,}
Rabi oscillations along one period $T_{\rm Rabi}$ for all spins of the processor.
We measure the spin-up probability of the nucleus $P_\Uparrow$ (electron $P_\uparrow$) as function of the coherent NMR (ESR)  drive duration.
\textbf{e,}
Phase coherence times measured for each spin via Ramsey (open) and Hahn-echo (filled) measurements.
\textbf{f,}
Single-qubit randomised-benchmarking (1Q-RB) results for each qubit showing average physical gate fidelities.
}
\label{fig:setup}
\end{figure*}

The connectivity of the nuclei and electrons both within each register and across registers is central to the operation of the 11-qubit atom processor (see Fig.~\ref{fig:setup}a).
Each spin register contains nuclei ($n_{1-4}$ and $n_{5-9}$) that are hyperfine coupled to a common electron ($e_{1}$ and $e_{2}$).
Importantly, these electrons are also exchange-coupled to each other enabling non-local connectivity across the registers (see Fig.~\ref{fig:setup}b). 
The strength of electron-exchange coupling, $J$, is tunable by the voltage detuning $\varepsilon$ across in-plane control gates (see Fig.~\ref{fig:setup}c and Supplementary Materials~I). 
Here, we operate in a weak exchange-coupled regime with $J\approx1.55$~MHz (see Fig.~\ref{fig:setup}c).
In this regime, the controlled rotations (CROT) on the electron are less susceptible to charge noise and not conditional on the nuclear spins in the other register~\cite{Kalra2014,He2019,Kranz2023high,Stemp2025}.
We note, that the CROT operation on the electron spin has the advantage of implementing a native multi-qubit Toffoli gate that is conditional on the nuclear spins.

The initial calibration of the 11-qubit atom processor requires the characterisation of $2^4+2^5=48$ electron-spin resonances (ESR) which is doubled to 96 in the presence of electron-exchange interaction.
Analysing the stability of the ESR peaks (see Supplementary Materials~II), we find that the frequencies within each register shift collectively.
Accordingly, we can implement an efficient recalibration protocol that scales linearly with the number of coupled spin registers.
By characterising the ESR frequency for a single 
reference configuration of the nuclear spins, we infer the exact positions of all other ESR transitions of the register from the frequency offsets of the initial calibration.
As a result, recalibrating all 96 ESR frequencies requires only two measurements\textemdash {\it i.e.} one per register.

The readout of an individual nuclear spin is performed via quantum non-demolition (QND) readout using the ancillary electron (see Supplementary Materials~III).
For nuclear spin initialisation, we combine this ESR-based approach with conditional NMR $\pi$ pulses (see Supplementary Materials~IV).
To maximise the certainty of the initialised state, we perform QND readout of the nuclear spin configuration of each register prior to each experiment and apply post-selection on the desired nuclear spin configuration.
The excellent contrast in Rabi oscillations (see Fig.~\ref{fig:setup}d) for all data qubits reflects the performance of the nuclear-spin readout and initialisation procedure.

The coherence times for both nuclear and electron spins are characterised via Ramsey and Hahn echo measurements (see Fig.~\ref{fig:setup}e).
For the nuclear spins, the phase coherence time extracted from Ramsey measurements, $T_2^\star$, ranges from 1 to 46 ms.
Refocusing with Hahn echo significantly extends such a phase coherence, $T_2^{\rm Hahn}$, to values between 3 and 660 ms.
We observe that the phase coherence of the data qubits is related to its hyperfine Stark coefficient (see Supplementary Material~V).
Accordingly, we note that deterministic atom placement will provide a way to improve coherence by tailoring the spin registers for smaller susceptibility to electric field fluctuations.
For the electrons $e_{1}$ and $e_{2}$, we measure similar phase coherence times of $T_2^\star\approx 20$~$\mu$s and $T_2^{\rm Hahn}\approx 350$~$\mu$s.
Overall, our investigations affirm the potential of refocusing techniques to significantly improve the performance of our 11-qubit atom processor.

Single-qubit randomised benchmarking (1Q-RB) reveals that all qubits except $n_4$ operate with gate fidelities above 99.9\% and as high as 99.99\% for $n5$ (see Supplementary Materials~VI for optimisation details). 
The excellent performance of our atom processor arises from a combination of long coherence times, fast gate operation, and minimal frequency drifts in both ESR and NMR (see Supplementary Materials~II and V).
These single-qubit metrics are on par with our recent results using a single spin register~\cite{Thorvaldson2024}, indicating consistency in atomic-scale fabrication.

\begin{figure}[h]
\includegraphics[width=0.47\textwidth]{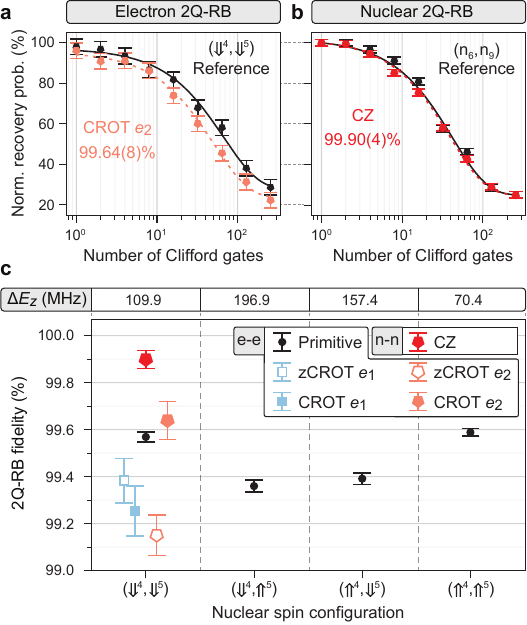}	
\caption{
\textbf{
High-fidelity two-qubit operation between nuclear (CZ) and electron (CROT) spins.
}
\textbf{a,}
Normalised two-qubit-randomised-benchmarking (2Q-RB) of the electron-electron CROT gate from the reference (black) and interleaved procedure (CROT $e_2$) showing the Clifford fidelity.
\textbf{b,}
Normalised 2Q-RB of the geometric CZ operation on the nuclear spin pair $n_6$ and $n_{9}$ from the reference (black) and interleaved procedure (CZ).
All other nuclear spins are initialised to down $\Downarrow$ in this experiment.
\textbf{c,}
Summary of the nuclear (CZ) and electron 2Q-RB (zCROT and CROT of $e_1$ and $e_2$) fidelities.
For the electron CROT gate, the primitive fidelities (reference for interleaved 2Q-RB) are also shown for different nuclear spin configurations with the corresponding frequency gap $\Delta E_z=|f_{\textrm{CROT e1}}-f_{\textrm{CROT e2}}|$ indicated at the top.
}
\label{fig:exchange}
\end{figure}

To now enable universal control of our 11-qubit atom processor, we establish a quantum link between the nuclear-spin registers via the exchange interaction of the electrons.
We first assess the performance of this link with interleaved two-qubit randomised-benchmarking (2Q-RB) of the electron CROT gate (see Methods).
Here, we calibrate the phase offsets of the CROT gates and implement a compensation protocol~\cite{Wu2024} (see Supplementary VI).
Figure~\ref{fig:exchange}a shows the reference and interleaved 2Q-RB data for $e_2$ when all nuclear spins are initialised to down ($\Downarrow\Downarrow\Downarrow\Downarrow$, $\Downarrow\Downarrow\Downarrow\Downarrow\Downarrow$) which we denote for simplicity as ($\Downarrow^4$, $\Downarrow^5$).
The extracted CROT gate fidelity of $99.64(8)$\% indicates beyond fault-tolerance performance of the electron-electron two-qubit gate.

According to Ref.~\cite{Kranz2023high}, the fidelity of the two-qubit CROT gate depends on the Larmor-frequency splitting $\Delta E_z=|f_{\textrm{CROT e1}}-f_{\textrm{CROT e2}}|$ which is defined by the nuclear-spin configuration.
In particular, when $\Delta E_z$ is similar to the exchange interaction strength $J$, the fidelity is lower due to hybridisation with the singlet-triplet eigenbasis.
By choosing small small exchange of $J=1.55$~MHz, we operate at a large $\Delta E_z/J$ ratio and obtain CROT gate fidelities >99\% across different nuclear-spin configurations as shown in  Fig.~\ref{fig:exchange}c.

A key task of the ancilla electron in our 14|15 platform is to entangle nuclear data qubits via a geometric CZ gate that is implemented via a $2\pi$-ESR rotation (2X gate)~\cite{Madzik2022,Thorvaldson2024}.
Figure \ref{fig:exchange}b shows interleaved 2Q-RB results for the nuclear CZ gate applied on two nuclear spins $n_6$ and $n_{9}$ on the 4P register giving a nuclear two-qubit-gate fidelity of 99.90(4)\%.
The nuclear CZ gate strongly outperforms the CROT gate and thus allows local multi-qubit operation on a spin register with high fidelity.

\begin{figure*}[t!]
\centering
\includegraphics{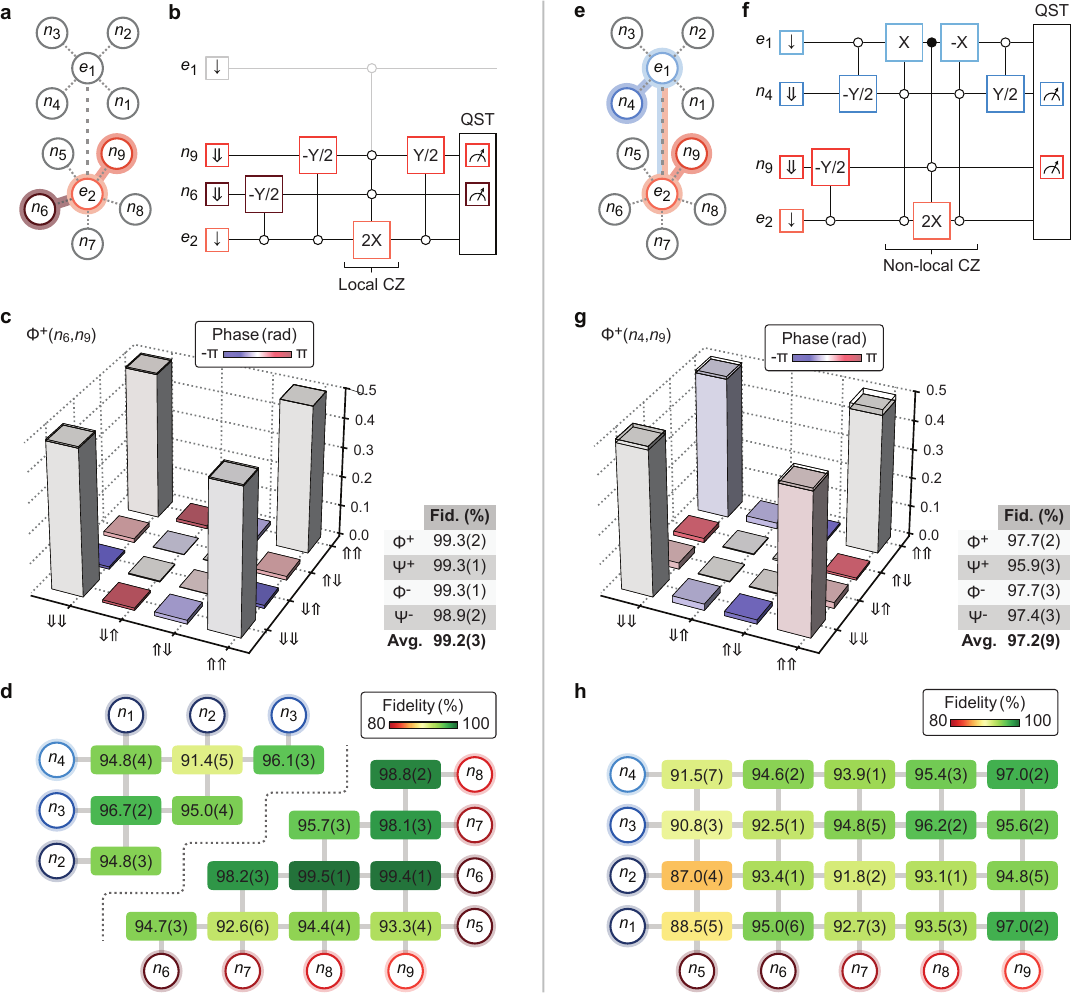}	
\caption{
\textbf{Bell-states within a register (left\textemdash local) and across registers (right\textemdash non-local).}
\textbf{a (e),}
Connectivity of a local (non-local) Bell state.
\textbf{b (f),}
Circuit for generation and measurement of a $\ket{\Phi^+}$ Bell state using local (non-local) CZ gate and quantum-state tomography (QST).
\textbf{c (g),}
Reconstructed density matrix for a local (non-local) $\ket{\Phi^+}$ Bell state.
The table shows the fidelities for all local (non-local) Bell states. 
Here, a complete set of 9 projections is used to reconstruct the density matrix.
\textbf{d (h),}
Generation fidelities of local (non-local) $\ket{\Phi^+}$ Bell state for all combinations of nuclear spins. 
As we use a reduced set of 3 projections, small deviations in the generation fidelities occur.
}
\label{fig:bell}
\end{figure*}

Before applying this electron-exchange-based link to entangle nuclear spins across the two registers, we first benchmark the generation of local Bell states within a single spin register.
As an example, we entangle the nuclear spins $n_6$ and $n_9$ of the 5P register via the electron $e_2$ (see schematic in Fig.~\ref{fig:bell}a).
An exemplary quantum circuit to prepare the Bell state is shown in Fig.~\ref{fig:bell}b which utilises this nuclear CZ gate to entangle the nuclear spin pair.
Accordingly, the four maximally entangled Bell states, $\ket{\Phi^{\pm}}=(\ket{\Downarrow\Downarrow}\pm\ket{\Uparrow\Uparrow})/\sqrt{2}$ and $\ket{\Psi^{\pm}}=(\ket{\Downarrow\Uparrow}\pm\ket{\Uparrow\Downarrow})/\sqrt{2}$ can be generated by adjusting the phase of the initial $-Y/2$ NMR pulses, by inverting their respective signs.
We perform quantum state tomography (QST) using a complete set of 9 projections (all combinations of X, Y and Z for the two data qubits) and reconstruct the corresponding density matrix (see Methods and Fig.~\ref{fig:bell}c).
Without removal of state-preparation and measurement (SPAM) errors, we obtain an average state fidelity of 99.2(3)\% for all Bell states (see Table in Fig.~\ref{fig:bell}c).
To characterise the local $\ket{\Phi^+}$ state across all nuclear-spin pairs from the two registers, we reconstruct the density matrix from a reduced set of three projections (XX, YY and ZZ).
This way, we can increase the measurement speed with minor reduction in accuracy.
Figure~\ref{fig:bell}d shows the local $\ket{\Phi^+}$ state fidelities for all local combinations of data qubits on the respective registers ranging from 91.4(5)\% to 99.5(1)\%.
To the best of our knowledge, the peak Bell-state fidelity surpassing 99\% represents the highest value reported in semiconductor devices to date.

As a next step, we now interconnect the spin registers and implement non-local Bell states over the electron-exchange-based link.
To demonstrate the approach, we entangle nuclear spin $n_4$ and $n_{9}$ via both electrons $e_1$ and $e_2$ (see connectivity in Fig.~\ref{fig:bell}e).
To implement the non-local CZ gate in the regime where $J\ll\Delta E_z$, we project the targeted nuclear state on the electron $e_1$ via $X$ gates ($\pi$ rotation) sandwiching the $2X$ operation on $e_2$ (see circuit in Fig.~\ref{fig:bell}f for the example of the $\ket{\Phi^+}$ state).
Again, we perform QST using a complete set of 9 projections to maximise measurement accuracy.
Figure~\ref{fig:bell}g shows the density matrix of the non-local Bell state $\ket{\Phi^+}$ with a table listing the extracted fidelities for $\ket{\Phi^+}$, $\ket{\Phi^-}$, $\ket{\Psi^+}$ and $\ket{\Psi^-}$ with an average of 97.2(9)\%.
We characterise the non-local $\ket{\Phi^+}$ state for all combinations of nuclear-spin pairs across the registers.
Figure~\ref{fig:bell}h shows the obtained state fidelities ranging from 87.0(4)\% to 97.0(2)\%.
The characterisation of non-local entanglement for all combinations of nuclear spins across the registers shows that our 11-qubit atom processor is capable of efficient all-to-all connectivity.

\begin{figure*}[!h]
\centering
\includegraphics{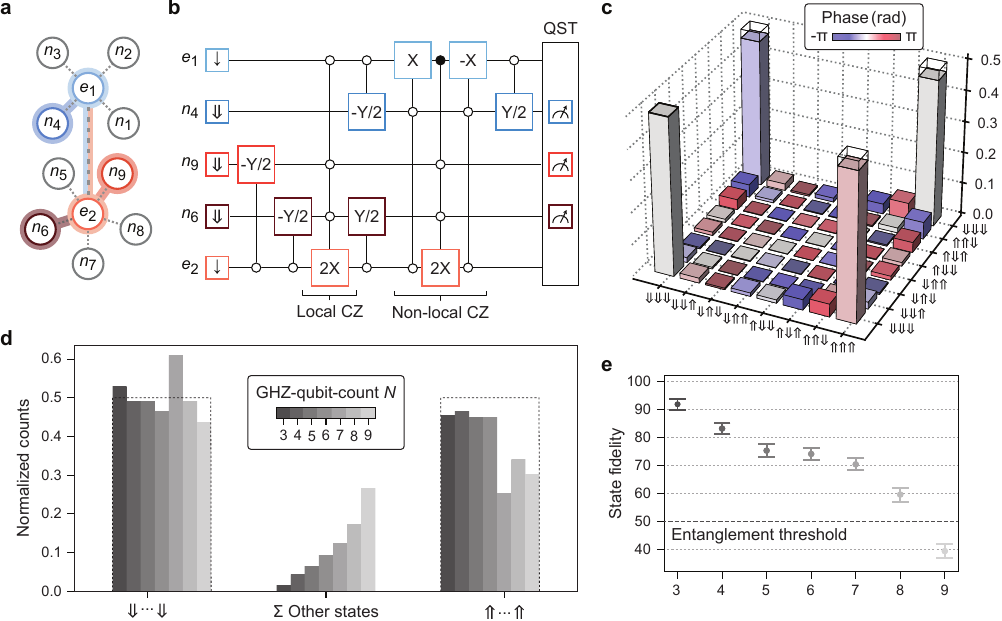}	
\caption{
\textbf{Non-local multi-qubit GHZ states.}
\textbf{a,}
Connectivity of the three-qubit-GHZ state comprising $n_4$, $n_6$ and $n_9$.
\textbf{b,}
Circuit for the generation and measurement of the three-qubit-GHZ state using the local and non-local CZ gate and quantum-state tomography (QST) via the ancilla qubits $e_1$ and $e_2$.
\textbf{c,}
Reconstructed density matrix for the GHZ state with $N=3$ entangled nuclear spins.
\textbf{d,}
Normalised QST counts in the $z$ projection\textemdash {\it i.e.} diagonal of density matrix\textemdash for GHZ states with increasing qubit count $N$.
The bars on the left (right) show matrix elements where all nuclear spins are down $\Downarrow...\Downarrow$ (up $\Uparrow...\Uparrow$).
All other elements with mixed states (with $\Downarrow$ and $\Uparrow$) are combined in the bars in the middle.
\textbf{e,}
Generation fidelity as function of the number of qubits $N$ in the GHZ state.
}
\label{fig:ghz}
\end{figure*}

The extension of entanglement over all data qubits is a critical benchmark for a quantum processor.
Accordingly, we investigate in the following non-local multi-qubit entanglement with an increasing number of nuclear spins.
First, we generate the Greenberger–Horne–Zeilinger (GHZ) state with three nuclear spins: $n_4$ on the 4P register and $n_6$ and $n_9$ on the 5P register (see Fig.~\ref{fig:ghz}a).
We implement a combination of local and non-local Bell states and concatenate the corresponding entanglement circuits as shown in Fig.~\ref{fig:ghz}b.
The density matrix shown in Fig.~\ref{fig:ghz}c is reconstructed from a full set of QST measurements.
Without SPAM removal, we report a GHZ state fidelity of 90.8(3)\%.

To prepare a GHZ state with more than 3 qubits, we now extend the circuit shown in Fig.~\ref{fig:ghz}b by adding the local entanglement sequence\textemdash NMR $-Y/2$, local ESR $2X$ and NMR $Y/2$\textemdash for each additional qubit.
For the 5P (4P) register we add these local entanglement operations before (after) the non-local CZ.
Because the number of tomography bases grows exponentially ($3^N$, where $N$ is the number of qubits), we use a reduced measurement strategy that requires only $N+1$ bases to estimate the state fidelity~\cite{Ghne2007, Moses2023}. 
Figure~\ref{fig:ghz}d shows the counts in the $z$ basis of GHZ states with an increasing number of entangled nuclear spins.
In the ideal GHZ state, measurement outcomes are equally distributed between the states in which all nuclear spins are either down ($\Downarrow$...$\Downarrow$) or up ($\Uparrow$...$\Uparrow$).
Increasing the number of qubits in the GHZ state ($N$) we observe a gradual increase in the probability of all other states, \textit{i.e.}, mixed combinations of $\Downarrow$ and $\Uparrow$.
The corresponding GHZ fidelities are plotted in Fig.~\ref{fig:ghz}e.
The 3-qubit GHZ fidelity is 92(2)\%, consistent with the value of 90.8(3)\% obtained from full quantum state tomography.
Since a fidelity above 50\% is sufficient to witness genuine $N$-qubit entanglement~\cite{Ghne2010}, the data demonstrates that entanglement is maintained for up to 8 nuclear spins.
Further performance improvements are anticipated by coherent control optimisation, frequency crosstalk mitigation and the incorporation of refocusing pulses.
Building on this progress, the present results demonstrate efficient connectivity across nuclear data qubits in our atom processor, representing an important step towards future implementations of quantum error correction on the 14|15 platform.

\section*{Conclusion}

The efficient connectivity of multi-nuclear spin registers within the 14|15 platform marks a major advancement of semiconductor spin qubits.
By coupling a 4P and a 5P register via electron-exchange interaction, we considerably exceeded the number of fully-interconnected qubits with respect to previous works in semiconductor devices~\cite{Madzik2022,Hendrickx2021,Philips2022,Zhang2024,Thorvaldson2024}.
While increasing the number of connected qubits, we have shown that physical-level benchmarks are maintained and some of them even improved, with two-qubit gate fidelities reaching 99.9\% for the first time in silicon qubits.
Systematic characterisation of the 11-qubit atom processor enabled the development of tailored calibration routines that scale linearly with additional registers.
By employing the electron spin on each of the two registers as an ancilla qubit, we implemented efficient single- and multi-qubit control for all nuclear spins.
This level of performance has allowed us to entangle every nuclear-spin pair within the 11-qubit system with Bell-state fidelities ranging from 91.4(5)\% to 99.5(1)\% within registers and from 87.0(4)\% to 97.0(2)\% across registers. 
We expanded the connectivity by preparing multi-qubit GHZ states across all data qubits and showed that entanglement is preserved for up to 8 nuclear spins.
Future work will focus on application of benchmarking methods such as gate-set~\cite{Nielsen2021} and non-Markovian tomography~\cite{White2020} to further optimise qubit control via pulse shaping, parallel drive execution and refined resonance frequency shift mitigation protocols~\cite{Undseth2023}.
By successfully introducing a coherent link across spin registers while maintaining excellent qubit performance, we demonstrate a key capability for future implementations in the 14|15 platform aimed at quantum error correction.
\section*{Methods}
\label{sec:methods}

{\bf Experimental setup.}

A single electron transistor (SET) serves as charge reservoir and sensor enabling spin readout of the electrons via spin-to-charge conversion.
Details to the basic operation of our atom processor are provided in Supplementary Material I.
The encapsulation is about 45 nm. 
On top of the chip, an antenna is horizontally offset from the dots by $\approx$300 nm \cite{hile2018addressable,Thorvaldson2024}.
It allows us to drive nuclear magnetic resonance (NMR) and electron spin resonance (ESR).
The experiment is performed in a cryogen-free dilution refrigerator at a base temperature of about 16 mK.
Spin polarisation is activated by a magnetic field $B\approx1.39$ T along the [110] crystal direction. 

{\bf Randomised benchmarking.}

For single-qubit randomised benchmarking (1Q-RB), we generate 10 variations of a random set of Cliffords up to 1024 gates.
Each Clifford gate is chosen from the one-qubit Clifford group containing 24 elements.
Using the Euler decomposition, we translate each Clifford to a single native $Y(\theta)$ rotation sandwiched between two virtual $Z(\theta)$ gates.
Since the latter operation is instantaneous due to a change of reference frame, the average number of primitive gates per Clifford is exactly one.
For each Clifford set, we take 200 (50) single-shot measurements for the electron (nuclei).
We perform qubit frequency recalibrations every 12 runs (equivalent to a few minutes time intervals).
We measure recovery probabilities $F_{\uparrow}(n)$ and $F_{\downarrow}(n)$ to both up and down states, and fit the data points with $F(n)=F_{\uparrow}(n)-F_{\downarrow}(n)$ with $F(n)=Ap^n$ where $n$ is the sequence length, $A$ is the factor containing SPAM errors and $p$ is the depolarizing strength.
The Clifford gate fidelity $F_C$, and hence the primitive gate fidelity $F_P$, is then extracted as $F_C=F_P=(1+p)/2$.
In all RB experiments, we calculate the error bars by bootstrapping re-sampling methods assuming a multinomial distribution~\cite{Huang2019, Wu2024}.

Similarly, for two-qubit randomised benchmarking (2Q-RB), we typically generate 20 variations of a random set of Cliffords up to 256 gates.
Each Clifford gate is chosen from the two-qubit Clifford group containing 11520 elements~\cite{Barends2014}.
For the electron, we employ the decomposition to CROT rotations as in Ref.~\cite{Huang2019}, where the average number of primitive gates, $\bar{n}$, is 2.57.
For nuclear spins, the native operations consist of a combination of $\pi/2$ NMR pulses for single-qubit rotations and $2\pi$ ESR pulses as CZ gates~\cite{Thorvaldson2024}.
Similar to 1Q-RB, we measure recovery probabilities to both $\uparrow\uparrow$ and $\downarrow\downarrow$ to account for SPAM errors.
To extract the polarizing strength, we fit $F(n)=F_{\uparrow\uparrow}(n) - F_{\downarrow\downarrow}$ with $F(n)=Ap^n$ as before.
The corresponding Clifford and primitive gate fidelities are $F_C=(1+3p)/4$ and $F_P=1-(1-F_C)/\bar{n}$, respectively.

For the interleaved 2Q-RB, we insert the target Clifford after each random gate, effectively doubling the sequence length.
We measure the recovery probabilities in the same manner as standard 2Q-RB, and extract the interleaved polarizing strength $p_i$.
Accordingly, we extract the interleaved gate fidelity via $F_i=(1+3p_i/p)/4$.
The standard deviation is calculated using the same bootstrapping re-sampling method and standard error propagation analysis.

{\bf Quantum state tomography.}

To perform quantum state tomography (QST) measurements, we add projection pulses for each qubit to the target basis $\{x,y,z\}$ prior to readout.
In particular, we apply $-Y/2$ ($X/2$) to project on $x$ ($y$).
Since $z$ is our native basis, there is no need for any extra rotations.

For Bell-state and GHZ-state generation, we merge the projection pulse with the last $Y/2$ rotation.
Accordingly, when projecting to $z$, the rotations cancel each other, and thus we remove them both.
For projections to $y$, we convert the sequence $Y/2+X/2$ into $-Z/2+Y/2$ according to the  Euler decomposition since the virtual $Z$ rotation which is implemented by a global phase shift does not require a physical pulse.

The full quantum state tomography is taken by projecting to all $3^N$ basis where $N$ is the number of qubits involved. 
We perform 2000 single-shot measurements per basis, and apply post-selection to ensure successful nuclear spin initialisation.
The density matrix is reconstructed by performing a constrained Gaussian linear least-squares fit to the tomography counts.
The standard deviation is then extracted from Monte Carlo bootstrapping re-sampling~\cite{Watson2018, Huang2019, Thorvaldson2024}.

\section*{Competing interests}

M.Y.S. is a director of the company Silicon Quantum Computing Pty Ltd. 
H.E., J.W.,  A.M.S., I.T., M.T.J, S.H.M., W.J.P., C.M., Y.H., H.B., S.K.G., Y. C., J.G.K., L.K. and M.Y.S. (all authors) declare equity interest in Silicon Quantum Computing Pty Ltd.

\section*{Author contributions}

J.W., H.E. and A.M.S. measured the device with the help of W.J.P. and C.M. under the supervision of L.K.;
I.T., Y.H. and S.K.G. provided theoretical support to the measurements.
M.T.J, S.H.M., and H.B. fabricated the device under the supervision of Y. C. and J.G.K.;
The manuscript was written by H.E. and J.W. with input from all authors.
L.K and M.Y.S. supervised the project.
These authors contributed equally: H.E. and J.W.
These authors jointly supervised this work: L.K. and M.Y.S.

\section*{Data availability}
The raw data used in this Article are available via Zenodo at \href{https://doi.org/10.5281/zenodo.15549984}{https://doi.org/10.5281/zenodo.15549984}

\section*{Code availability}
The code used to analyse the data and produce the figures in this Article is available via Zenodo at \href{https://doi.org/10.5281/zenodo.15549984}{https://doi.org/10.5281/zenodo.15549984}

\newpage
\newpage$\,$\newpage
\label{references}
\bibliography{references}

\end{document}


\preprint{APS/123-QED}

\title{
{\Large SUPPLEMENTARY MATERIALS}\\
An 11-qubit atom processor in silicon
}

\author{Hermann Edlbauer}
    \affiliation{\sqc}
    \affiliation{\rm These authors contributed equally to this work.}
\author{Junliang Wang}
    \affiliation{\sqc}
    \affiliation{\rm These authors contributed equally to this work.}
\author{A. M. Saffat-Ee Huq}
    \affiliation{\sqc}
\author{Ian Thorvaldson}
    \affiliation{\sqc}
\author{Michael T. Jones}
    \affiliation{\sqc}
\author{Saiful Haque Misha}
    \affiliation{\sqc}
\author{William J. Pappas}
    \affiliation{\sqc}
\author{Christian M. Moehle}
    \affiliation{\sqc}
\author{Yu‐Ling Hsueh}
    \affiliation{\sqc}
\author{Henric Bornemann}
    \affiliation{\sqc}
\author{Samuel K. Gorman}
    \affiliation{\sqc}
\author{Yousun Chung}
    \affiliation{\sqc}
\author{Joris G. Keizer}
    \affiliation{\sqc}
\author{Ludwik Kranz}
    \affiliation{\sqc}
    \affiliation{\rm These authors jointly supervised this work.}
\author{Michelle Y. Simmons}
    \affiliation{\sqc}
    \affiliation{\rm These authors jointly supervised this work.}
    \affiliation{\rm Corresponding author: \href{mailto: michelle.simmons@sqc.com.au}{michelle.simmons@sqc.com.au}}
\maketitle

\onecolumngrid
\tableofcontents

\newpage
\addtocounter{page}{-1}

\section{Basic operation of the 11-qubit atom processor}
\label{suppl:basic_operation}

\begin{figure}[h!]
\centering
\includegraphics[trim={0 0cm 0cm 0cm},clip]{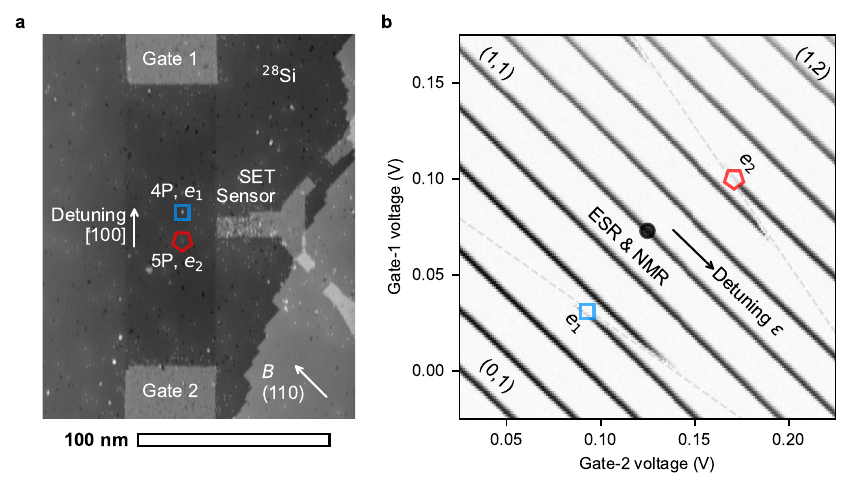}	
\caption{
\textbf{Basic operation of the 11-qubit atom processor via detuning gates.}
\textbf{a,}
Scanning tunneling micrograph showing the core of the atom processor with the detuning gates 1 and 2 and the single electron transistor (SET).
The antenna that drives ESR and NMR is located on top of the silicon chip.
\textbf{b,}
Gatemap showing the antenna-operation point (ESR \& NMR) and working points for initialisation and readout of the electrons ($e_1$ and $e_2$).
The grey, dashed lines indicate the separation between the charge configurations (N,M) with integer numbers N and M indicating the number of electrons in the 4P and 5P register.
The arrow indicates the direction of voltage detuning $\varepsilon$ of gate 1 and 2 to shift the antenna-operation point to configurations with stronger electron exchange.
}
\label{sfig:stm_gatemap}
\end{figure}

We operate the spin registers via two gates (gate 1 and 2), a single-electron transistor (SET) and an on-chip antenna.
Figure~\ref{sfig:stm_gatemap}a shows the hydrogen lithography mask for the gates and the SET after incorporation of the spin registers.
Besides its reservoir functionality, the SET serves as charge-sensor via a radio-frequency (RF) reflectometry setup enabling ramped single-shot spin readout \cite{Keith2022ramped}.
Gate 1 and 2 control the number of electrons on each register that are loaded from a single-electron transistor (SET) located 17(1) nm away from the registers (center to edge). 
Figure~\ref{sfig:stm_gatemap}b shows the SET signal (in arbitrary units) measured over the (1,1) charge where one electron is loaded on each of the spin registers.
To initialise and readout the spins of the electrons, we move to the breaks of the Coulomb peaks indicated as $e_1$ and $e_2$.
Between spin initialisation and readout, we bring the gate voltages into an intermediate position where we drive nuclear-magnetic-resonance (NMR) and electron-spin-resonance (ESR) via a broadband antenna, which is located on top of the chip at a horizontal distance of $\approx$300~nm \cite{hile2018addressable,Thorvaldson2024}.

\newpage
\section{Characterisation and recalibration of electron-spin resonances (ESR)}
\label{suppl:stability}

\subsection{Frequency stability of ESR}
\label{suppl:esr_stability}

To track the resonant ESR frequencies conditional on a given reference and target nuclear spin state, we employ the quantum circuit shown in Fig.~\ref{sfig:esr_stability}a.
The schematic shows the protocol by the example of electron $e_1$ and a subsystem of corresponding nuclear spins $n_1$ and $n_2$.
To minimise the chance for frequency jumps or drifts, we interleave ESR measurements for a reference and target state.
Prior to the reference measurement we initialise all electron and nuclear spins of the system to the down state ($\downarrow\Downarrow^4$,$\downarrow\Downarrow^5$).
After the reference measurement, we transition into the target state via a set of NMR $\pi$ rotations ($X$ gates) on the specific nuclear spins\textemdash that is only a single $X$ gate on $n_2$ in the present example.
High-precision ESR measurements are achieved by coherent ESR drive of $n$ periods of $\pi$ ($X^n$).
Here we have chosen $n=9$ rotations to balance accuracy versus measurement duration.
As we perform a fine frequency sweep of the two signals around the reference ($f_0$) and target ($f_1$) frequency, be obtain two precise spectra and thus the ESR offset.

Investigating the ESR stability, we find different results for the 4P and 5P register:
Figure~\ref{sfig:esr_stability}b shows time evolutions of the ESR frequencies $f_{\rm ref.}$ of each register conditional to all nuclear spins being in the down-state ($\Downarrow^4$,$\Downarrow^5$).
The corresponding histograms are shown in Fig.~\ref{sfig:esr_stability}c.
Remarkably, the 4P register is stable for the entire tracking period (up to 15 hours) with a standard deviation $\sigma=0.90(5)$~kHz.
For the 5P register, on the other hand, we observe distinct frequencies indicating the presence of two-level systems (TLS) with an extent of 10~kHz and 45~kHz.
Since these jumps are only observed for one register, we rule out the possibility of electrostatic fluctuations which should affect both registers equally.
We performed ionised NMR experiments \cite{Holmes2024}, and observed no signal from Si$^{29}$.
For the fidelity of qubit manipulations, the broadening of the Rabi frequency $f_{\rm Rabi}\approx430$~kHz renders this 10~kHz shift negligible.
The 45-kHz jumps are less frequent and occur over tens of minutes or longer.
By performing recalibration every 5 to 10 minutes for sensitive experiments such as Bell State tomography, we are able to mitigate the effect of these slow fluctuations.
Based on these results, we speculate that the stability of the ESR frequency is either determined by the specific geometry of the nuclear spin registers or growth imperfections.
Accordingly, we anticipate an increased ESR stability with further advances in deterministic-incorporation techniques.

\subsection{ESR-calibration protocol for multi-nuclear spin registers}
\label{suppl:esr_recal}

The presence of any TLS would require frequent recalibration of ESR frequencies.
Since the number of ESR frequencies is $2^k$ for each multi-nuclear spin register, where $k$ is the number phosphorus atoms, there are in total 48 ESR ($2^4$+$2^5$) frequencies to keep track of when the electrons are initialised to $\downarrow$.
A typical recalibration measurement for a single frequency takes approximately 1 minute. Consequently, performing a full calibration of the system could take up to 50 minutes. Such a procedure is impractical, as the earliest calibrated frequencies may already have drifted by the time the process is complete.

\begin{figure}[p!]
\centering
\includegraphics[trim={0 0cm 0cm 0cm},clip]{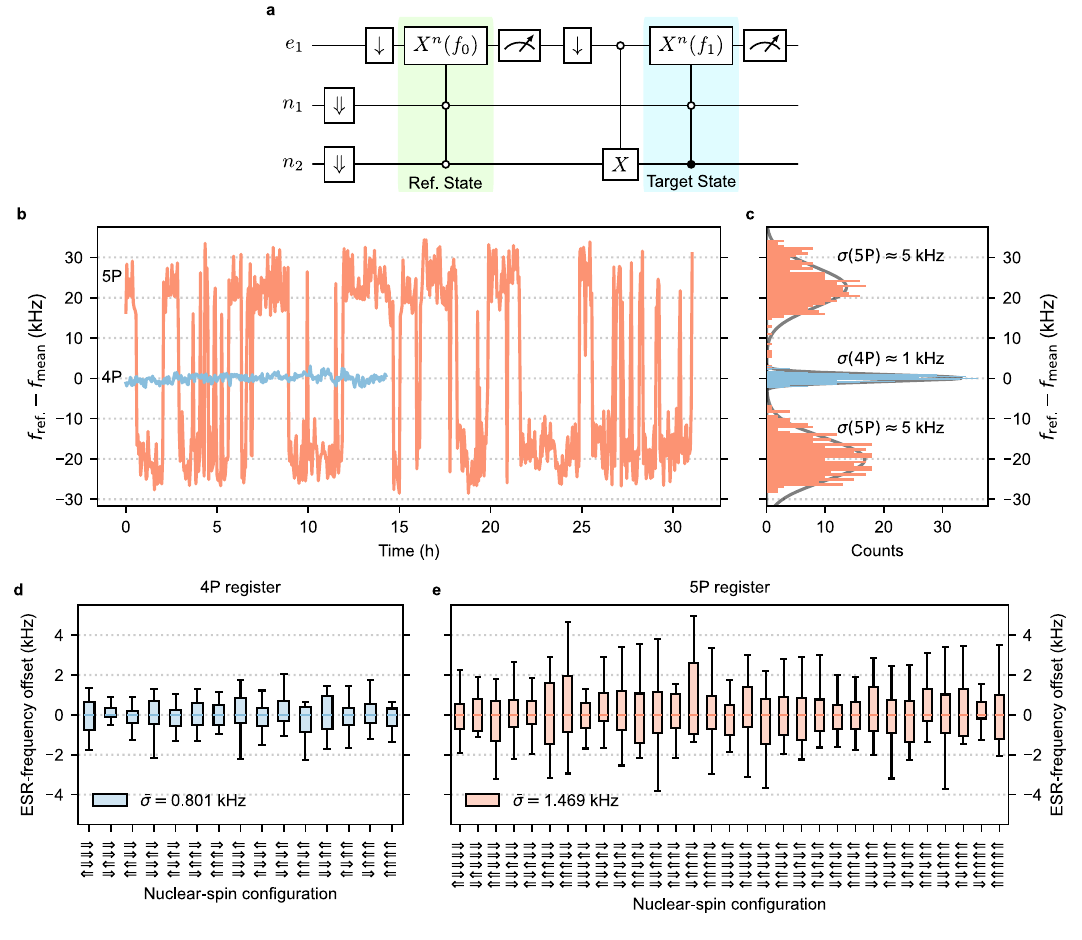}	
\caption{
\textbf{Stability of the ESR frequencies over time.}
\textbf{a,}
Circuit diagram for tracking the relative offset between an ESR transition for a target nuclear spin and a reference nuclear spin state.
$X^n$ indicates the number $n$ of $\pi$ rotations on the electron.
$f_0$ is the frequency parameter to scan across the reference ESR peaks ($f_{\rm ref.}$) where all nuclear spins are in the down state ($\Downarrow^4$,$\Downarrow^5$).
$f_1$ is the frequency parameter to measure a targeted ESR peak ($f_{\rm target}$) in any other nuclear configuration.
$f_0$ and $f_1$ are swept in parallel to extract the resonant ESR frequencies $f_{\rm ref.}$ and $f_{\rm target}$ relative to each other.
\textbf{b,}
Evolution of the reference ESR transition in which all nuclear spins $\Downarrow$ for 4P (blue) and 5P (orange) over several hours.
To compare the traces from the 4P and 5P register, we subtract $f_{\rm mean}$ from the data.
Here, $n=9$ rotations are employed.
\textbf{c,}
Histograms of $f_{\rm ref.}$ from \textbf{b}. 
\textbf{d (e),}
Boxplots of the frequency offset $f_{\rm target}-f_{\rm ref.}$ distribution for all possible nuclear spin states in the 4P (5P) register with the mean standard deviation, $\bar{\sigma}$.
For comparison, the values are centered at the median value.
}
\label{sfig:esr_stability}
\end{figure}

To overcome this challenge, we analyze the correlations in the ESR fluctuations for different nuclear spin-configurations.
Figure~\ref{sfig:esr_stability}d and~\ref{sfig:esr_stability}e show the statistical distribution of the ESR frequency offset for all nuclear spin configurations referenced to the ($\Downarrow^4$,$\Downarrow^5$) ESR.
Since the fluctuations are similar for all states, we find that the ESR frequencies shift collectively.
The mean standard deviation of the 4P register, $\bar{\sigma}_{\rm 4P}\approx0.8$~kHz, is consistent with the extracted deviation of $1$~kHz from tracking $f_{\rm ref.}$.
For the 5P system, all frequency offsets exhibit deviations less than 10~kHz.
Since $\bar{\sigma}_{\rm 5P}\approx1.5$~kHz is much lower than 45~kHz which is the extent of the largest fluctuation on the 5P register (compare Fig. ~\ref{sfig:esr_stability}bc), the TLS jumps occur simultaneously for all ESR resonances.
Therefore, by calibrating the reference frequency $f_{\rm ref.}$ we can update all drive frequencies employing the characterised frequency offsets\textemdash proving a scalable ESR-recalibration protocol for multi-nuclear spin registers.

\newpage

\subsection{Frequency stability of electron-exchange coupling $J$}
\label{suppl:exch_stability}

\begin{figure}[h!]
\centering
\includegraphics[trim={0 0cm 0cm 0cm},clip]{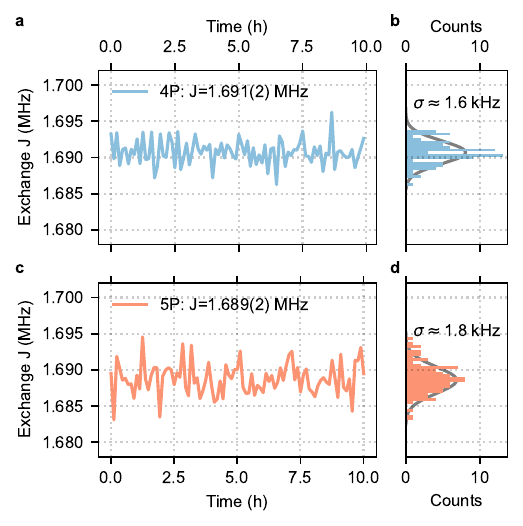}	
\caption{
\textbf{Stability of the exchange interaction, $J$.}
\textbf{a (c),}
Evolution of the exchange strength $J=f_{\rm CROT}-f_{\rm zCROT}$ from 4P (5P) over 10 hours.
The employed quantum circuit is a modified version of~\ref{sfig:esr_stability}a where the reference and target transitions are the zCROT and CROT operations, respectively.
Here, all nuclear spins are initialised to $\Downarrow$.
\textbf{b (d),}
Histogram of $J$ from \textbf{a} (\textbf{c}).
}
\label{sfig:exchange_stability}
\end{figure}

The stability measurement of the exchange interaction $J$ is similar to the protocol shown in Fig.~\ref{sfig:esr_stability}a.
In the reference measurement, we have all spins initialised to the down state: ($\downarrow\Downarrow^4$,$\downarrow\Downarrow^5$).
By having the control-electron in the $\downarrow$ state, the corresponding ESR measurement is thus probing the zCROT branch.
To measure electron-exchange coupling, the target measurement is modified accordingly:
Instead of changing the nuclear configuration, we invert the control electron to the spin-up-state $\uparrow$ via an ESR $\pi$ pulse and thus target the corresponding CROT branch.
As we repeatedly measure the frequency gap between the zCROT and CROT branches, we can characterise the fluctuations of the electron-exchange coupling $J$.
Figure~\ref{sfig:exchange_stability} shows the results of such stability measurement for the electrons $e_1$ and $e_2$ on the 4P and 5P registers.
Over 10 hours, the fluctuations of $J$ remain below 1.8~kHz, indicating that any TLS jumps do not affect the exchange interaction.
Accordingly, we can simply recalibrate zCROT only and assign the drive frequency for CROT using the frequency gap ($J=1.690$~MHz in this configuration).
Compared to the full calibration protocol, we thus half the number of recalibration measurements for electron two-qubit-gate operations.

\newpage

\section{Quantum-non-demolition (QND) nuclear-spin readout}
\label{suppl:nuclear_read}

\begin{figure}[b!]
\centering
\includegraphics[trim={0 0cm 0cm 0cm},clip]{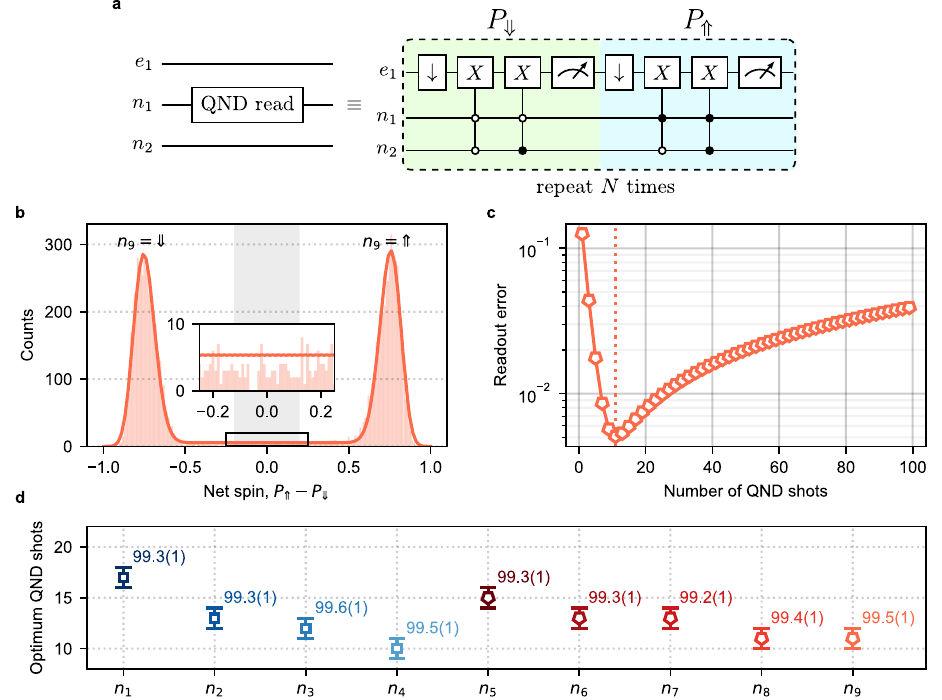}	
\caption{
\textbf{Quantum-non-demolition (QND) readout of the nuclear spins.}
\textbf{a,}
Exemplary QND-readout circuit of nuclear spin $n_1$ in the reduced subsystem with dependency on $n_2$ only.
\textbf{b,}
Histogram of the net spin, $P_{\Uparrow}-P_{\Downarrow}$, from $N=100$ QND shots for qubit $n_{9}$.
The solid line is the fit using a Markov chain model.
The grey shaded region (typically, $\pm 0.2$) highlights the range at which net spin values are rejected in post-selection to remove uncertain spin classifications.
The inset shows the histogram baseline.
\textbf{c,}
The readout error extracted from \textbf{b} where the dotted line indicates the optimum number of QND shots for highest readout fidelity. 
\textbf{d,}
Optimum number of QND shots for all nuclear spins and their corresponding readout fidelity in percent (\%).}
\label{sfig:nndr}
\end{figure}

To measure the state of the nuclear spin, we perform quantum non-demolition (QND) readout \cite{Pla2013, Thorvaldson2024}.
Using the electron spin as an ancilla qubit, we map the target nuclear spin state to the electron spin via a sequence of conditional ESR pulses.
Figure~\ref{sfig:nndr}a shows an example of a QND-readout circuit for a subsystem of $n_1$ and $n_2$ of the 4P register.
To account for variations in electron-spin readout visibility, we perform two sequences of $2^{k-1}$ $\pi$-ESR rotations (X gate) that are conditional on $n_1=\Downarrow$ and $n_1=\Uparrow$ and unconditional on the other nuclear spins. Here, $k$ is the total number of nuclear spins in the register.
In the present example, $k=2$ with $n_2$ being the only other nuclear spin considered.
As the ESR drives are sandwiched by an electron $\downarrow$-initialisation and readout, we project the nuclear spin on the electron state.
Repeating the cycle $N$ times we obtain $P_{\Uparrow}$ and $P_{\Downarrow}$.
Then, we classify the nuclear spin being $\Uparrow$ as $\Delta P = P_{\Uparrow} - P_{\Downarrow} \ge 0$.

In contrast to previous works~\cite{Thorvaldson2024}, we use fast coherent rotations instead of slow adiabatic inversion pulses to drive the ESR transitions.
In this way, we can speed-up readout cycles and reduce the probability of nuclear spin-flip events during the QND readout.
A prerequisite to implementing such coherent drive process is to have all ESR resonance frequencies well calibrated
\textemdash see Supplementary Material~\ref{suppl:stability}.

To characterise nuclear-spin readout, we perform 1000 repetitions of the circuit with $N=100$ read shots.
An exemplary histogram of $\Delta P$ from such an experiment is shown in Fig.~\ref{sfig:nndr}b.
By fitting the two distinct peaks with a Markov-chain model~\cite{Thorvaldson2024}, we reproduce the finite background (inset), which is caused by nuclear spin-flip events during the QND readout cycles.
Setting the threshold range appropriately (grey region), we can assess the nuclear-spin-readout fidelity without bias from spin-flip events.
Employing the Markov-chain model, we extract the readout error as a function of the number of QND shots as shown in Fig.~\ref{sfig:nndr}c.
Despite an electron-spin readout fidelity of $\approx$75\%, we can increase increase the number of QND shots to reduce nuclear-spin-readout error.
Beyond a certain point, the likelihood of nuclear spin flips becomes significant, leading to a degradation in overall fidelity.
The optimum number of shots for QND readout of each nuclear spin is shown in Fig.~\ref{sfig:nndr}d.
Remarkably, we find that we can achieve nuclear-spin-readout fidelity above the fault-tolerant threshold of 99\% for all data qubits. 

The protocol to readout a specific set of states for all the nuclear spins is almost identical to the circuit that is required to obtain a single, specific nuclear spin state as shown in Fig.~\ref{sfig:nndr}a.
Below, we show a similar example for a readout circuit of both $n_1$ and $n_2$ with the target state being $\Downarrow\Downarrow$:
\begin{equation*}
\begin{quantikz}[row sep=0.4cm]
\lstick{$e_1$} & \qw & \qw\\
\lstick{$n_1$} & \gate[wires=2]{\shortstack{QND read \\ state: $\Downarrow\Downarrow$}}  & \qw \\
\lstick{$n_2$} & \qw & \qw
\end{quantikz}
\equiv
\begin{quantikz}[row sep=0.4cm, column sep=0.2cm]
\lstick{$e_1$} & \gate{\downarrow} & \qw & \gate{X} \gategroup[wires=3, steps=3, style={dashed, rounded corners, draw=black}, label style={label position=above, yshift=+0.1cm}]{$P_{\neq{\Downarrow\Downarrow}}$} & \gate{X} & \gate{X} & \qw & \meter{} & \gate{\downarrow}   & \qw & \gate{X}  \gategroup[wires=3, steps=1, style={dashed, rounded corners, draw=black}, label style={label position=above, yshift=+0.1cm}]{$P_{\Downarrow\Downarrow}$}& \qw& \meter{}  & \qw\\
\lstick{$n_1$}      & \qw      & \qw  & \octrl{-1}    & \ctrl{-1} & \ctrl{-1}     & \qw & \qw   & \qw  & \qw & \octrl{-1} & \qw & \qw& \qw\\
\lstick{$n_2$}      & \qw    & \qw    & \ctrl{-1}     & \octrl{-1} & \ctrl{-1}   & \qw & \qw   & \qw   & \qw & \octrl{-1} & \qw & \qw& \qw
\end{quantikz}
\end{equation*}
The first half of the circuit ($P_{\neq{\Downarrow\Downarrow}}$) maps all the nuclear states that are different to the target spin state $\Downarrow\Downarrow$ to the electron spin.
Subsequently, we realise the mapping of the target state and calculate the equivalent net value $P_{\rm net} = P_{\Downarrow\Downarrow}-P_{\neq{\Downarrow\Downarrow}}$
We then assign the nuclear state as $P_{\rm net}\geq0$.
Note that this protocol offers a speed advantage over sequential nuclear-spin readout, since it only extracts a single binary outcome indicating if all nuclear spins in the register occupy a specific state.
In short, instead of performing QND readout measurements for each of the $k$ nuclear spins, only one sequence is executed.

\section{Optimisation of nuclear spin initialisation}
\label{suppl:nuclear_init}

The ability to initialise all nuclear spins from a random state to a certain initial state (such as all $\Downarrow$) is essential for the operation of our quantum processor.
We follow a protocol used in previous works~\cite{Thorvaldson2024}
based on a process called electron state transfer (EST)~\cite{Waldherr2014}, followed by a verification QND nuclear-state 
readout to increase initialisation fidelity by post-selecting on the desired target state.
An example of an EST circuit initializing $n_1$ in the subsystem with $n_2$ is shown below:
\begin{equation*}
\begin{quantikz}[row sep=0.4cm]
\lstick{$e_1$} & \qw & \qw \\
\lstick{$n_1$} & \gate{\Downarrow} & \qw \\
\lstick{$n_2$} & \qw & \qw
\end{quantikz}
\equiv
\begin{quantikz}[row sep=0.4cm, column sep=0.2cm]
\lstick{$e_1$} & \gate{\downarrow}  &  \gate{X} & \gate{X} & \ctrl{1} & \qw\\
\lstick{$n_1$} & \qw & \ctrl{-1} & \ctrl{-1}  & \gate{X} & \qw\\
\lstick{$n_2$} & \qw & \octrl{-1} & \ctrl{-1} & \qw & \qw
\end{quantikz}
\end{equation*}

The central idea is to rotate the nuclear spin conditional on the electron spin.
For instance, to initialise a given nuclear spin $n_1$ to $\Downarrow$, we first prepare the electron spin in the $\downarrow$ state with high-fidelity.
Subsequently, we rotate the electron to $\uparrow$ using ESR conditional to $\ket{n_1}=\ket{\Uparrow}$, but unconditional to the other nuclear spins.
If $k$ nuclear spins are present in the multi-nuclear spin register, $2^{k-1}$ ESR frequencies are needed to be driven in order to probe the state of a specific data qubit.
After the state of the nuclear spin has been projected on the electron state,
a final NMR $\pi$ rotation conditional to $\ket{e}=\ket{\uparrow}$ is applied to prepare the nuclear spin to $\Downarrow$.
Note that if $\ket{n_1}=\ket{\Downarrow}$ from the beginning, the sequence has no effect on the nuclear spin.
To initialise all nuclear spins to $\Downarrow$, we perform such a EST sequence for each spin sequentially\textemdash with up to 3 repetitions if high fidelity is required.
To reach an initialisation fidelity above 99\%, we perform a QND nuclear state readout after the EST sequence and discard undesired states via post-selection.

\newpage
\section{Frequency stability of nuclear magnetic resonances (NMR)}
\label{suppl:nmr_stability}

\begin{figure}[h!]
\centering
\includegraphics[trim={0 0cm 0cm 0cm},clip]{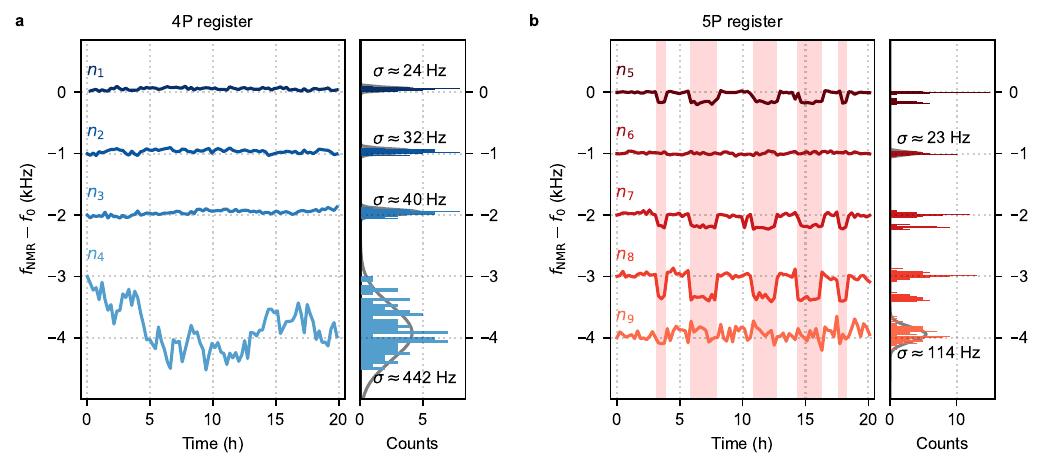}	
\caption{
\textbf{Stability of the NMR frequencies}
\textbf{a (b),}
Evolution of resonant NMR frequencies over time for nuclear spins in the 4P (5P) register.
The first data point $f_0$ is subtracted to highlight relative changes and the traces are vertically shifted for clarity.
Red shaded regions show correlated jumps between $n_{5}$, $n_{7}$ and $n_{8}$.
}
\label{sfig:nmr_stability}
\end{figure}

\begin{figure}[h!]
\centering
\includegraphics[trim={0 0cm 0cm 0cm},clip]{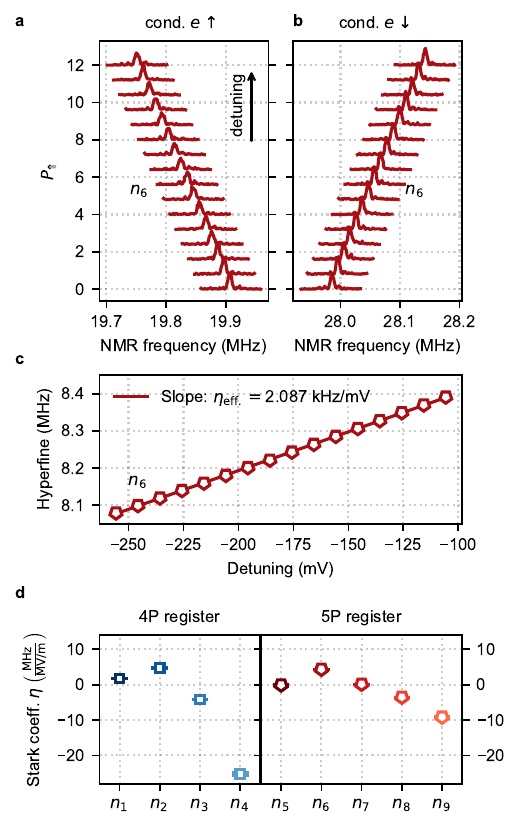}	
\caption{
\textbf{Hyperfine Stark shift\textemdash method and values for all nuclear spins.}
\textbf{a (b),}
Coherent NMR spectrum of $n_{6}$ conditional on the electron spin up-state $\uparrow$ (down-state $\downarrow$) as a function of detuning, $\varepsilon$. 
Traces are vertically shifted by 0.8 for clarity.
\textbf{c,}
The value of the hyperfine coupling in MHz extracted from NMR peak locations in \textbf{a} and \textbf{b}.
The slope of the linear fit (solid line) represents the effective Stark coefficient, $\eta_{\rm eff.}$.
\textbf{d,}
Stark coefficient, $\eta=\eta_{\rm eff.} d_{\rm gate}/\alpha$, obtained by normalizing $\eta_{\rm eff.}$ with the electric field employing the distance between the gates $G_1$ and $G_2$ $d_{\rm gate}=146.6$ nm and the V-to-eV conversion factor $\alpha=0.07$.
}
\label{sfig:stark}
\end{figure}

We also assess fluctuations of the NMR frequencies by repeatedly measuring NMR spectra.
We first initialise the system in the ($\downarrow\Downarrow^4$,$\downarrow\Downarrow^5$) configuration.
Subsequently, we apply a specific NMR drive before finally reading the nuclear spin via QND readout.
All nuclear spins in the 4P register shows fluctuations below 50~Hz (see Fig.~\ref{sfig:nmr_stability}a), except for $n_4$ which exhibits a non-negligible drift (up to 1.3~kHz) over 10 hours.
We find that the observed NMR stability correlates with the hyperfine Stark coefficient, $\eta$, of the nuclear spins (see Fig.~\ref{sfig:stark}d).
Owing to large $\eta$ value of $n_4$ which is $\approx-25$~MHz/(MV/m), this nuclear spin is more susceptible to charge noise.
We can mitigate the impact of such a strong frequency drift by more frequent recalibrations of $n_4$.
On the contrary, the 5P register shows no slow drifts (see Fig.~\ref{sfig:nmr_stability}b).
However, a striking feature of this data is the correlated jump of the nuclear spins $n_5$, $n_7$ and $n_{8}$ (red shaded regions).
Note that, since these frequency jumps are less than 500~Hz in magnitude, any changes in hyperfine values are not sufficient to explain the 45~kHz jumps observed in the ESR frequency tracking data shown in Fig.~\ref{sfig:esr_stability}b.
We remark that for $n_1$, $n_2$, $n_3$ of the 4P register and $n_6$ of the 5P register, we find remarkable NMR stability with fluctuations below 50~Hz.

\newpage$\,$\newpage
\section{Optimisation of gate operations}
\label{suppl:gate_optimization}

\subsection{Compensation of frequency shifts induced by microwave drive}
\label{suppl:esr_nmr_warming}

\begin{figure}[h!]
\centering
\includegraphics[trim={0 0cm 0cm 0cm},clip]{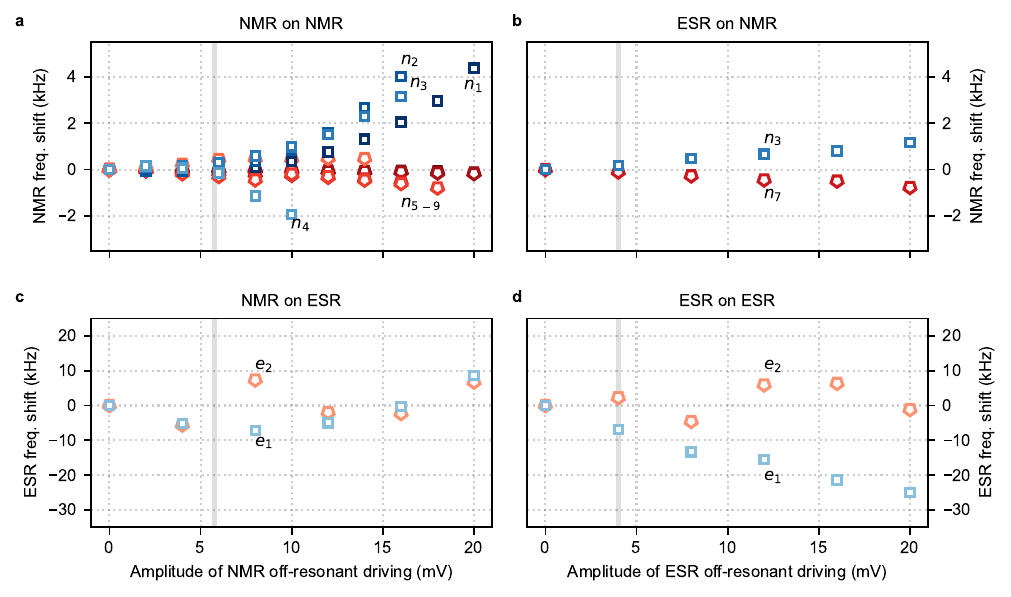}	
\caption{
\textbf{Shifts in electron and nuclear spin-qubit frequencies with off-resonant microwave drive.}
\textbf{a,}
Frequency-shift of a nuclear-magnetic-resonance (NMR) peaks by off-resonant NMR drive  (``NMR on NMR'') with increasing amplitude.
The frequency shifts are recorded via nuclear Ramsey experiments with off-resonant NMR drive at $f_w=50$~MHz during the idling time.
An initial microwave burst of 500~$\mu$s at $f_w$ is applied before the Ramsey pulse and a waiting time of 2~ms before the readout.
The off-resonant drive amplitude is the applied peak-to-peak voltage at room temperature without taking into account attenuation.
\textbf{b,}
Frequency-shift of NMR peaks by an off-resonant electron-spin-resonance (ESR), namely ``ESR on NMR", with increasing amplitude.
Similar experiment to \textbf{a}, but using instead an off-resonant ESR drive at $f_w=38.86$~GHz during the idling time.
Here, we have only characterised one nuclear spin from each register, $n_4$ (4P) and $n_7$ (5P), as the impact on our measurements is comparably small in this case.
\textbf{c,}
Frequency-shift of ESR peaks by an off-resonant NMR drive (``NMR on ESR'') with increasing amplitude.
Similar experiment than in \textbf{a}, but performing an electron Ramsey sequence with NMR off-resonance driving during idling.
\textbf{d,}
Frequency-shift of ESR peaks by an off-resonant ESR drive (``ESR on ESR'') with increasing amplitude.
For the present experiment, we set the off-resonant NMR and ESR amplitudes to 5.75~mV and 4~mV (shaded grey).
}
\label{sfig:warming}
\end{figure}

Microwave-induced frequency shifts during spin qubit operation in semiconductor devices have been repeatedly reported in the literature~\cite{Philips2022,Undseth2023,Tanttu2024}.
We investigate this effect in our multi-nuclear spin register by running Ramsey-like experiments with the nuclear- and electron-spin qubits in which we replace the idling between the $\pi/2$-pulses with off-resonant NMR (ESR) driving at 50 MHz (38.86 GHz).
The period of the resulting Ramsey oscillations thus reflects the frequency shift that is introduced by the intermediate off-resonant drive.
Figure~\ref{sfig:warming} shows the corresponding frequency shift as function of the amplitude of the off-resonant driving power.
The four panels show different cases where an ESR or NMR Ramsey pulse is executed with either ESR or NMR off-resonant driving.
For example the panel ``ESR on NMR'' shows the results of an NMR-Ramsey experiment sandwiched with an off-resonant ESR drive of varying amplitude in mV.

We first focus on the influence of the NMR drive amplitude on the NMR resonance frequency\textemdash see Fig.~\ref{sfig:warming}a.
Here we observe quadratic changes in the NMR frequency as we increase the power of the off-resonant NMR signal.
In contrast to the work by Undseth \textit{et al.}~\cite{Undseth2023}, we observe no sign of saturation over the range investigated.
A striking feature of the ``NMR on NMR'' data is the different behavior for each of the registers: 
The nuclear spins of the 5P register show significantly smaller shifts.
We suspect that the difference stems from the specific geometry of the phosphorus atoms withing the multi-nuclear spin registers.
Accordingly, we anticipate that advances in deterministic phosphorus atom placement will enable us to reduce the susceptibility to NMR driving.
For all the other cases ``ESR on NMR'', ``NMR on ESR'' and ``ESR on ESR''\textemdash see Fig.~\ref{sfig:warming}b-d\textemdash, we observe smaller impact on qubit operation.
For electron spins $e_1$ and $e_2$, we observe a non-monotonous shifts in ESR frequency what poses a challenge for consistent operation.
In general we find that the extent of the ESR and NMR shifts increase with ESR or NMR driving power.

Overall, the data shows that both ESR and NMR qubit frequencies change during any idling time if the total microwave input power is not kept constant during circuit operations.
To compensate for such coherent errors, we equalise the effective input power applied at any time by calibrating the electron (nuclear) qubit frequencies with parallel NMR (ESR) drive at an off-resonant frequency, $f_{\rm off}^{n/e}$, as shown in the circuit below:
\begin{equation*}
\begin{quantikz}[slice all,remove end slices=1,slice
titles= , slice style=black]
\lstick{$e$} & \gate{X} & \octrl{1} & \gate{\rm Idle} & \qw \\
\lstick{$n$} & \octrl{-1} & \gate{X} & \gate{\rm Idle} & \qw
\end{quantikz}
\equiv
\begin{quantikz}[slice all,remove end slices=1,slice
titles= , slice style=black]
\lstick{ESR} & \gate{f_{\Downarrow}} & \gate{f_{\rm off}^{e}} & \gate{f_{\rm off}^{e}} & \\
\lstick{NMR} & \gate{f_{\rm off}^{n}} & \gate{f_{\downarrow}} & \gate{f_{\rm off}^{n}} & \qw
\end{quantikz}
\end{equation*}
We thus establish a fixed budget for the antenna drive that covers the effective load for both ESR and NMR.
In this work, we set $f_{\rm off}^{n} = 50$ MHz and $f_{\rm off}^{e}=38.86$ GHz.
This approach ensures consistent input power during antenna operation with any partial or complete idling periods.

\newpage
\subsection{Optimisation of NMR Rabi frequencies}
\label{suppl:nmr_rabi_tuning}

The amplitude of the NMR pulses determines the corresponding Rabi frequency $f_{\rm Rabi}$ and is thus important for efficient operation.
Faster rotations make it possible to carry out more operations before error sources have an impact.
On the other hand, stronger NMR drive causes shifts of NMR and ESR frequencies\textemdash see Fig.~\ref{sfig:warming}.
Accordingly, the optimal choice of the NMR drive power needs to balance these aspects.
Here, we use the fidelity of 1Q-RB experiments as metric to find the optimal value of $f_{\rm Rabi}$.

\begin{figure}[h!]
\centering
\includegraphics[trim={0 0cm 0cm 0cm},clip]{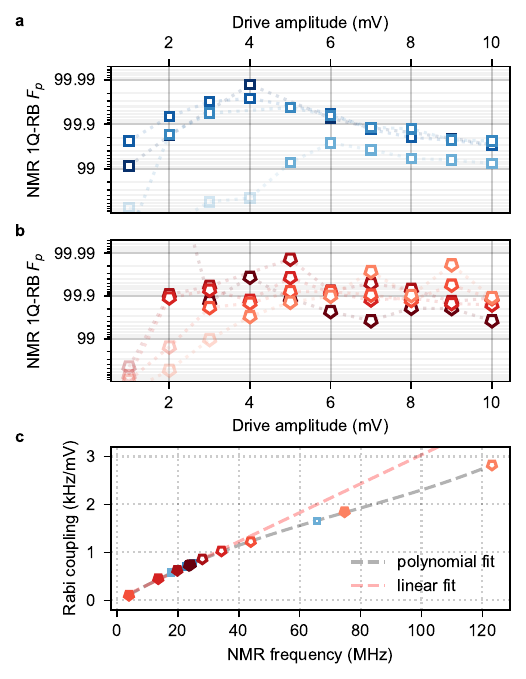}	
\caption{
\textbf{Optimisation of nuclear-spin drives via single-qubit randomised benchmarking (1Q-RB).}
\textbf{a (b),}
Primitive 1Q-RB fidelity as a function of the room-temperature drive amplitude for nuclear spins in the 4P (5P) register.
Data points above $99\%$ are highlighted.
\textbf{c,}
Rabi coupling, $f_{\rm Rabi}/A_{\rm drive}$, for coherent NMR frequencies conditional to electron down-state $\downarrow$ (open symbols) and up-state $\uparrow$ (closed symbols).
The data points are extracted using Rabi frequencies $f_{\rm Rabi}$ from \textbf{a} and \textbf{b}.
Grey dashed line is a fit using a polynomial function of degree 3.
To highlight the linearity between $f_{\rm Rabi}$ and $f_{\rm NMR}$, data points below 40~MHz are fitted with a linear function (red dashed line).
}
\label{sfig:nmr_rb}
\end{figure}

For the 4P register, the resulting 1Q-RB data\textemdash see Fig.~\ref{sfig:nmr_rb}a\textemdash shows 
the effect of NMR power on the gate fidelity for all nuclear spins.
Above a drive amplitude of 4 mV, where maximal fidelity $>$99.95\% is observed, the Larmor frequency starts to shift with a concomitant drop in performance.
This observation is not true for $n_4$ which shows an optimal performance at 6 mV.
For the range of drive amplitudes investigated, we observe a linear dependency on the Rabi frequency and no change on $T_2^*$ (data not shown).
For the 5P register, no drop in 1Q-RB fidelity is observed as the NMR-drive amplitude is increased\textemdash see Fig.~\ref{sfig:nmr_rb}b.
The data is consistent with the shifts in nuclear spin-qubit frequencies induced by off-resonant NMR drive which saturates at a smaller magnitude\textemdash compare pentagon markers in Fig.~\ref{sfig:warming}a.

The NMR amplitudes for maximal 1Q-RB fidelity only remain a good choice if the absorbed power, which for NMR is proportional to the ratio of the Rabi and Larmor frequency $f_{\rm Rabi}/f_{\rm NMR}$, is constant.
Otherwise, if multiple NMR pulses are employed, the power budget varies during execution owing to the different shifts in the Larmor frequency.
We plot the Rabi coupling versus the NMR-Larmor frequency in Fig.~\ref{sfig:nmr_rb}c.
We find a deviation from a linear trend for drive frequencies beyond 30 MHz.
For NMR pulses driven beyond this threshold, we therefore need to increase the drive amplitude to keep the ratio of  $f_{\rm Rabi}/f_{\rm NMR}$ constant.
The corresponding increase in drive amplitude is extracted via a polynomial fit of degree 3 (dashed line) to the data.
A list of typical NMR drive parameters after such adjustments are listed in Table~\ref{stab:nmr}.
The Table is sorted by NMR-Larmor frequency.
The off-resonant NMR compensation pulse\textemdash for idling \textemdash is set to $f_{\rm NMR}=50$~MHz and $A_{\rm NMR}=5.421$~mV to achieve similar power absorption to the parameters listed. 
We note that this procedure cannot be generalised for ESR owing to a different relation of the absorbed power to the Rabi and Larmor frequency.

\renewcommand{\thetable}{S\Roman{table}}%

\begin{table}[h!]
\begin{tabular}{cccc}
\textbf{\#Qubit} & \textbf{$f_{\rm NMR}$ (MHz)} & \textbf{$f_{\rm Rabi}$ (kHz)} & \textbf{$A_{\rm NMR}$ (mV)} \\
$n_{5}$          & 24.221                       & 3.538                         & 4.789                       \\
$n_{1}$          & 24.450                       & 3.582                         & 4.726                       \\
$n_{2}$          & 24.534                       & 3.594                         & 4.725                       \\
$n_{3}$          & 24.985                       & 3.661                         & 4.750                       \\
$n_{6}$          & 28.110                       & 4.118                         & 4.783                       \\
$n_{7}$          & 34.324                       & 5.029                         & 4.952                       \\
$n_{8}$         & 43.902                       & 6.432                         & 5.327                       \\
$n_{4}$          & 65.412                       & 9.584                         & 5.895                       \\
$n_{9}$         & 123.026                      & 18.025                        & 6.393                      
\end{tabular}
\caption{
\textbf{Optimised NMR-drive parameters.}
To ensure the same power absorption, we keep the ratio $f_{\rm Rabi}/f_{\rm NMR}=1.46\times10^{-4}$ constant.
}
\label{stab:nmr}
\end{table}

\subsection{Optimisation of the electron-exchange-based CROT operation}
\label{suppl:2q_rb_ee}

To link multi-nuclear spin registers we operate an exchange-based CROT gate~\cite{Kalra2014,Kranz2023high} at low exchange coupling $J\sim1.5-1.7$ MHz, \textit{i.e.,} much smaller than the Larmor-frequency splitting of the CROT resonances of $e_1$ and $e_2$ of $\Delta E_z\approx109.9$ MHz.
In this regime, where $J\ll\Delta E_z$,
the CROT operation is less susceptible to charge noise and not conditional on the nuclear spin states of the other multi-nuclear spin register.

\begin{figure}[h!]
\centering
\includegraphics[trim={0 0cm 0cm 0cm},clip]{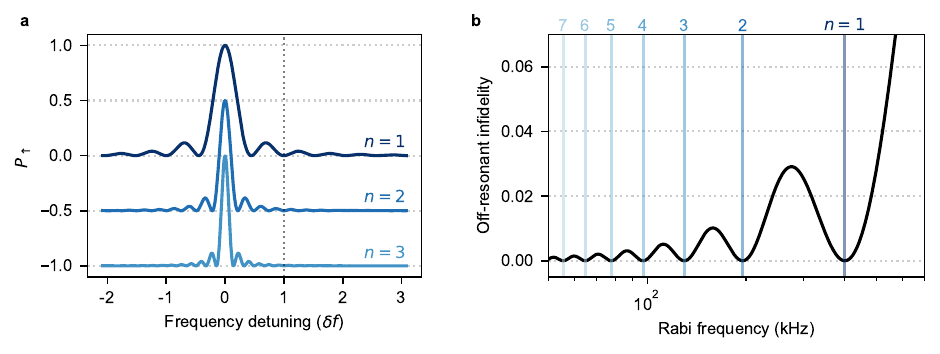}	
\caption{
\textbf{Optimisation of the Rabi frequency for high-fidelity CROT operation.}
\textbf{a,}
Analytical spin-flip probability for $\pi$ rotations for various $f_{\rm Rabi}(n)=\delta f / \sqrt{16n^2-1}$ where $n\in[1,2,3]$.
The traces are vertically shifted for better visualisation.
Here, for exchange operations, the CROT resonance frequency is located at $\delta f=J$\textemdash see dotted line.
\textbf{b,}
Infidelity, $1-P_{\uparrow}$, at a frequency detuning of $\delta f$ as function of Rabi frequency. The minima correspond to optimal Rabi frequencies $f_{\rm Rabi}(n)$ for $\pi$ rotations (vertical lines) where off-resonant drive is minimised at $\delta f$.}
\label{sfig:best_rabi}
\end{figure}

To further optimise the fidelity of the CROT operation, we can maximise its driving speed given by the Rabi frequency $f_{\rm Rabi}$.
Such an increase in Rabi speed, however, comes with power broadening since for $f_{\rm Rabi}\gtrsim J/4$ off-resonant drive of the other exchange resonance zCROT becomes significant and causes population transfer to the neighboring resonant peak.
This effect can be  particularly strong if no pulse shaping is applied (as in this work).
We can, however, mitigate the off-resonance drive if we choose the Rabi frequency such that the neighboring zCROT peak lies within a node of the power spectrum.
The spin-flip probability ($P_{s}$) of an antenna drive without any pulse modulation ({\it i.e.} boxcar window), follows the square of a sinc function:
\begin{equation}
    P_{s}=\frac{\Omega^2}{\Omega^2+\Delta^2}\sin^2{\left(\frac{t\sqrt{\Omega^2+\Delta^2}}{2}\right)}
\end{equation}
where $\Omega=2\pi f_{\rm Rabi}$ is the angular Rabi frequency, $\Delta = 2\pi \delta\!f$ is the angular frequency offset with respect to the Larmor frequency and $t$ is the duration of the pulse.
This function has local minima for a specific set of Rabi frequencies:
\begin{equation}
    f_{\text{Rabi}}(n)=\sqrt{\frac{n^2}{t^2}-\delta f^2}
\end{equation}
where $n$ is an integer number.
Optimizing $f_{\rm Rabi}$ for $\pi/2$ operations, we can minimise any off-resonant driving as shown in Fig.~\ref{sfig:best_rabi}.
For a $\pi/2$ rotation\textemdash with the duration $t = 1/(4\cdot f_{Rabi})$\textemdash, we obtain:
\begin{equation}
    f_{\text{Rabi}}(n)=\delta f/\sqrt{16n^2-1}.
\label{seq:frabi}
\end{equation}
To fulfill this condition and obtain a maximum Rabi frequency, we set $n=1$ for the discussed zCROT off-resonant drive of a CROT operation where $\delta\!f=J\approx1.55$~MHz.
Accordingly, we can minimise the off-resonant drive with a maximal speed at $f_{\rm Rabi}(n=1)\approx400$~kHz.

\begin{figure}[h!]
\centering
\includegraphics[trim={0 0cm 0cm 0cm},clip]{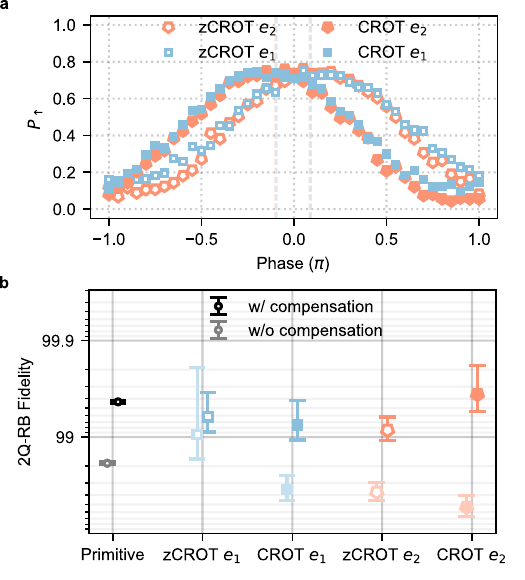}	
\caption{
\textbf{Optimisation of electron two-qubit randomised benchmarking (2Q-RB).}
\textbf{a,}
Phase shifts of coherent CROT operations following calibration sequences from Ref.~\cite{Wu2024}:
We execute a Ramsey-type experiment with $\pi/2$-zCROT (CROT) pulses sandwiching the complementary $\pi$-CROT (zCROT) pulse under test with the other electron being correspondingly initialised to $\downarrow$ and $\uparrow$.
The measurement is executed by sweeping the phase of the second $\pi/2$ pulse.
\textbf{b,}
e-e 2Q primitive (black) and interleaved RB fidelities (colors) with (solid) and without (semi-transparent) phase error compensation.
}
\label{sfig:2q_rb_ee}
\end{figure}

For the electron 2Q-RB data shown in Fig.~2 of the main paper, we also compensate for controlled-phase errors following the protocol defined by Ref.~\cite{Wu2024}.
The data obtained from measurements of phase-shift characterisation for the CROT operation are shown in Fig.~\ref{sfig:2q_rb_ee}a.
Applying the correction protocol, we observe a significant improvement in the standard and interleaved 2Q-RB as shown in Fig.~\ref{sfig:2q_rb_ee}b with all fidelity benchmarks above 99\%.
We remark however that by employing the electron spins as ancillary qubits, \textit{i.e.,} as an entanglement mechanism between the nuclear spins, such phase errors have negligible impact, and thus no corrections are needed.

\newpage

\newpage$\,$\newpage
\label{references}
\bibliography{references}